\documentclass[12pt]{article}
\pdfoutput=1

\usepackage{amsmath, amsfonts, amssymb}
\usepackage{graphicx}
\usepackage{psfrag}
\usepackage[usenames,dvipsnames,svgnames,table]{xcolor}
\usepackage{enumerate}
\usepackage{jheppub}
\usepackage{soul}
\usepackage{subfig}

\newcommand{\dd}{\textrm{d}}
\newcommand{\e}{\textrm{e}}

\newcommand{\w}{\wedge}

\newcommand{\SU}{\mathop{\rm SU}}

\newcommand{\be}{\begin{equation}}
\newcommand{\ee}{\end{equation}}
\newcommand{\bea}{\begin{eqnarray}}
\newcommand{\eea}{\end{eqnarray}}

\newcommand{\lp}{\left(}
\newcommand{\rp}{\right)}

\begin{document}
\title{Remarks on scale separation in flux vacua}
\author{F.F.~Gautason, M.~Schillo, T.~Van Riet and M.~Williams}

\emailAdd{ffg@fys.kuleuven.be, marjorie@itf.fys.kuleuven.be, thomas.vanriet@fys.kuleuven.be, m.williams@fys.kuleuven.b}

\affiliation{Instituut voor Theoretische Fysica, K.U. Leuven,\\
Celestijnenlaan 200D B-3001 Leuven, Belgium}

\abstract{We argue that the Maldacena-Nu{\~n}ez no-go theorem excluding Minkowski and de Sitter vacua in flux compactifications can be extended to exclude anti-de Sitter (AdS) vacua for which the Kaluza-Klein scale is parametrically smaller than the AdS length scale. As a practical application of this observation we demonstrate that the mechanism to resolve O6 singularities in massive type  IIA at the classical level is likely not to occur in AdS compactifications with scale separation. We furthermore remark that a compactification to four observable dimensions  implies a large cosmological hierarchy.}

\maketitle

\section{Introduction}
In any theory with extra dimensions, one is faced with the problem of stabilizing the extra dimensions at unobservable length scales.  In string theory, fluxes provide a mechanism for stabilizing compact directions, however assuring that these compactifications are genuinely four-dimensional requires that the masses of the Kaluza-Klein (KK) modes are out of the reach of four-dimensional observers.  It is believed that string theory provides a vast \emph{landscape} of vacua which exhibit this separation of scales.\footnote{A good overview can be found in \cite{Denef:2008wq, Grana:2005jc, Douglas:2006es}. Critical remarks about the mathematical evidence for a landscape can be found in \cite{Banks:2012hx}.}  This landscape, and a mechanism to populate it, is frequently employed to argue for an anthropic selection of the cosmological constant.

More concretely, a solution is scale separated if the KK-scale, $M_{\rm KK}$, can be decoupled from the scale of the cosmological constant, $M_{\Lambda}$:

\be\label{scalesep}
\frac{M_{\Lambda}}{M_{\rm KK}} \ll 1\,.
\ee
For example, Minkowski vacua are automatically scale separated once the extra dimensions are compact, since $M_{\Lambda}=0$. However, in the most straightforward flux compactifications it is usually difficult (or impossible) to achieve this separation of scales.

There is a close connection between scale separation (\ref{scalesep}) and the cosmological constant problem, which is often stated as $M_{\Lambda} \ll M_{\rm Planck}$. The appearance of the Planck scale in this inequality originates from the cut-off scale of a quantum field theory; usually the cut-off is taken to be the Planck mass since new physics must enter at that scale. However, if new physics enters before the Planck scale, it is more reasonable to state the cosmological constant problem in terms of this scale:
\be\label{ccproblem2}
\frac{M_{\Lambda}}{M_{\text{NewPhysics}}} \ll 1\,.
\ee
Clearly the KK-scale will introduce new physics in any theory with extra dimensions, therefore it is safe to treat $M_{\rm KK}$ as $M_{\text{NewPhysics}}$. It therefore seems that the string landscape is by definition a landscape of vacua with a cosmological constant problem.

Loosely speaking, the flux vacua of the landscape can be divided into two classes: first, quantum corrected no-scale vacua and second, tree-level AdS vacua. The first class is based on the classical four-dimensional Minkowski solutions of \cite{Dasgupta:1999ss, Giddings:2001yu}, which have moduli that are stabilized by perturbative and non-perturbative effects resulting in AdS vacua \cite{Kachru:2003aw, Balasubramanian:2005zx}, or perhaps even de Sitter vacua \cite{Westphal:2006tn}. The second class of vacua have all moduli stabilised at the ten-dimensional supergravity level and have only been found in massive type IIA supergravity with intersecting O6 planes \cite{Behrndt:2004km,Derendinger:2004jn,DeWolfe:2005uu, Acharya:2006ne,Caviezel:2008ik})\footnote{In type IIB similar constructions can be made but they are less understood \cite{Caviezel:2009tu, Petrini:2013ika}.}. 

These two constructions have one property in common: some flux charges are cancelled by orientifold planes. One of the main points of this paper is to emphasize that this is no coincidence and that no scale separated vacua can be found in supergravity without the introduction of negative tension objects. The rough idea is as follows, the Maldacena-Nu{\~n}ez (MN) no-go theorem \cite{Maldacena:2000mw} (first pointed out in \cite{Gibbons:1984kp,deWit:1986xg}), shows that, at the level of ten-dimensional supergravity, flux compactifications can only lead to Minkowski or de Sitter vacua if negative tension objects (such as orientifolds) are present. Although the string landscape consists of many AdS vacua, they all exhibit a hierarchy of scales in order to be perceived as four-dimensional. Our expectation is that any mechanism that is required to get $\Lambda \ge 0$ is also at work for negative but small $\Lambda$.  In other words: \emph{the same assumptions in the MN theorem that exclude Minkowski and de Sitter vacua can be used to exclude AdS vacua that are genuinely four-dimensional}. 

As an application of this idea we comment on a mechanism for resolving orientifold singularities found in \cite{Saracco:2012wc, Saracco:2013aoa}. The mechanism arose from a study of AdS flux vacua with scale separation within classical ten-dimensional supergravity. Using the findings in this note, we can argue that the classical resolution most likely does not occur when the AdS vacua feature scale separation.

\section{What is required for separation of scales?} \label{sec:MN}

As discussed in the introduction, a realistic compactification of a theory with extra dimensions requires a hierarchy 
of scales. The mass of the lightest massive Kaluza-Klein mode must exceed that set by the cosmological constant, $M_\text{KK}\gg M_\Lambda$. In order to verify this hierarchy for a given compactification, we must 
calculate both $M_\text{KK}$ and the cosmological constant $\Lambda$. Although it is often easy to determine $\Lambda$, the calculation of $M_\text{KK}$ requires detailed knowledge of the geometry of the internal manifold, $\mathcal{M}$. One must find the lowest-lying non-zero eigenvalue 
of the Laplacian on $\mathcal{M}$, but this is usually not feasible. 

In order to set the stage, we write the $D$-dimensional metric as a warped product of a maximally symmetric 4-dimensional metric and
$d$ extra dimensions,
\be \label{Mlam}
ds^2 = \e^{2 A(y)} \tilde{g}_{\mu \nu} dx^\mu dx^\nu + g_{mn}dy^m dy^n~.
\ee
The 4-dimensional metric  is $g_{\mu\nu}= \e^{2 A(y)} \tilde g_{\mu\nu}$, where $\mu,\nu=0,\dots,3$, and $g_{mn}$ is the 
internal $d$-dimensional metric of the compact manifold $\mathcal{M}$, with $m,n=1,\dots,d$. The full $D$-dimensional metric is denoted 
by $g_{MN}$ where $M,N=0,\dots,D-1$.  Since we take $\tilde g_{\mu\nu}$ to be maximally symmetric, the scale of the cosmological constant can be written simply in terms of the constant Ricci scalar of the unwarped metric:
\be 
M_{\Lambda}^2 = \tilde R_4~.
\ee
This unwarped curvature is related to the full 4-dimensional curvature via:
\be
\e^{4A} R_4= \e^{2 A} \tilde{R}_4 - \Box_d( \e^{4A})~.
\ee
Furthermore, in any warped compactification, the observed value of the 4-dimensional Planck mass is:
\be \label{Mplanck}
M_\text{Pl}^2 = M_D^{d+2}\int\sqrt{g_d}~\e^{2A}~\dd^d y~.
\ee 

In this paper we estimate $M_\text{KK}$ by the integrated 
curvature,
\be\label{estimate}
M_\text{KK}^{2-d}\sim \int \sqrt{g_d}~\e^{4A}~ R_d~ \dd^dy~,
\ee
where $R_d$ is the curvature scalar of the internal manifold.  This is partly inspired by dimensional analysis and also by a Theorem of Lichnerowicz \cite{Lich} which states that a compact manifold 
with a bounded Ricci curvature 
\be
R_d \ge (d-1)k>0~,
\ee 
where $k>0$ is some constant, has a lower bound on the first non-zero eigenvalue of the Laplacian, 
$\lambda_1\ge kd$. 
This is applicable to our case since we will see that the internal manifolds are positively curved.  Our goal in this section is to find a bound on the
ratio $M_\text{KK}/M_\Lambda$ in a broad class of flux compactifications of 10-- and 11--dimensional supergravity,
thereby measuring separation of scales in these compactifications.
The weak point in our argument will be the estimate above; our results will not hold for compactifications in which $M_\text{KK}$
greatly differs from \eqref{estimate}. 

Now that $M_{\Lambda}$ and $M_{\rm KK}$ have been defined in terms of curvature scalars, we can use the Einstein equations to write them in terms of the matter content:
\bea
R_4 &=& \frac{d-2}{d+2}T_4 - \frac{4}{d+2} T_d~,\label{externalcurv}\\
R_d &=& -\frac{d}{d+2}T_4 + \frac{2}{d+2}T_d~,\label{internalcurv}
\eea
where 
where 
\[
R_4=g^{\mu\nu}R_{\mu\nu}~,\quad T_4=g^{\mu\nu}T_{\mu\nu}~,\quad R_d = g^{mn}R_{mn}~,\quad T_d = g^{mn}T_{mn}~.
\]
In \cite{Gibbons:1984kp,deWit:1986xg} and later in \cite{Maldacena:2000mw} it was realized that for theories with neither positive potentials nor negative tension sources, all terms on the right
hand side of eq.~\eqref{externalcurv} are negative definite, thus, ruling out de Sitter vauca. However, they did not rule out the possibility that $R_4$ is very small and negative, giving rise to a scale separated vacuum; this is exactly the possibility we are investigating.

The equations (\ref{externalcurv}--\ref{internalcurv}) play an important role in measuring scale separation. The observationally relevant quantity is $M_{\rm Pl}^2 M_{\Lambda}^2$, so we compare this to $M_{\rm KK}^4$ to obtain a dimensionless measure of scale separation.  Then, we find that it is convenient to bound the right hand side of:
\be \label{ineqM}
\frac{M_\text{KK}^{4}}{M_\text{Pl}^2M_\Lambda^2} \ll \frac{M_\text{KK}^{4}}{M_\text{Pl}^2M_\Lambda^2} \left(\frac{M_D}{M_\text{KK}}\right)^{d+2} ~.
\ee
Combining \eqref{Mlam}-\eqref{estimate}, the ratio of integrated curvature scales relates 
directly to the right-hand-side of \eqref{ineqM}:
\be\label{theratio}
\frac{M_\text{KK}^{4}}{M_\text{Pl}^2M_\Lambda^2} \left(\frac{M_D}{M_\text{KK}}\right)^{d+2} \sim \left|\frac{\int \sqrt{g_d}~\e^{4A}~R_d~\dd^dy}{\int \sqrt{g_d}~\e^{2A}~\tilde{R}_4~\dd^dy}\right|
=\left|\frac{\int \sqrt{-g}~R_d~\dd^DX}{\int \sqrt{-g}~R_4~\dd^DX}\right|\equiv\left|\frac{\int R_d}{\int R_4}\right| ~.
\ee
We conclude that any scale separated vacuum must satisfy:
\be\label{crazynumber}
\left|\frac{\int R_d}{\int R_4}\right| \gg 1~.
\ee
If we are interested in AdS vacua possessing the same level of scale separation as our vacuum, we can estimate a lower bound of the left hand side of \eqref{ineqM}.  A reasonable lower bound of the KK scale is  $M_\text{KK}\sim \text{TeV}$ \cite{Chatrchyan:2013lca}, and using the observed value of $\Lambda$, we see that the ratio is $10^{60}$. 

Using equations (\ref{externalcurv}--\ref{internalcurv}) and other equations of motion we will show that in order to obtain
this kind of hierarchy the same ingredients must be present as are required for a de Sitter or Minkowski compactification, \emph{i.e.} orientifold planes or higher derivative corrections \cite{Douglas:2010rt}.
To do so we will work under almost the same assumptions as in \cite{Maldacena:2000mw}.  First we specialize to 10- or 11-dimensional supergravity and exclude all higher-derivative corrections.  Second, we assume that
spacetime is a maximally symmetric four-dimensional manifold, while the remaining
$d$-dimensional manifold is compact.  Strictly speaking, \cite{Maldacena:2000mw} also applies to non-compact 
manifolds where the warp factor is allowed to go asymptotically to 0, but we will not consider these cases.  

We warm up by looking for scale separation in 11-dimensional supergravity, simply because it has fewer ingredients.  After making the strategy clear in M-theory we will proceed to the 10-dimensional calculation where we also include the effect of D-branes.  Our derivation is inspired by the derivation in \cite{Petrini:2013ika}, which applies to the special case of unwarped solutions in IIA/IIB with constant dilaton.

\subsection{Scale separation in 11 dimensions}
The bosonic part of the eleven-dimensional supergravity action is particularly simple:
\be \label{Maction}
S_{11} = M_{11}^9 \int d^{11}X \sqrt{-g} \lp R - {1\over 2} |F_4|^2 \rp - \frac16 M_{11}^9\int F_4\w F_4\w A_3~,
\ee
with 
\[
|F_n|^2 = \frac{1}{n!}F_{M_1M_2\cdots M_n}F^{M_1M_2\cdots M_n}~, \qquad F_4=\dd A_3~.
\]
Calculating the energy momentum tensor and plugging in eqs. \eqref{externalcurv} and \eqref{internalcurv} yields

\bea
R_4 &=& - {4\over3}|F_{4}|^2 - {8\over3} |F_{7}|^2~,\\
R_7 &=&  {5\over3} \, |F_{4}|^2  + {7\over3} \, |F_{7}|^2~.
\eea
In these expressions we only make use of magnetic fluxes, \emph{i.e.}~field strengths with legs entirely along the internal
space. All external, or electric, fluxes that appear throughout the calculation are replaced by their magnetic duals, \emph{e.g.} $F_7
= \star_{11} F_4$.

We recognise that $R_4 \le 0$ as we expect from Maldacena-Nu{\~n}ez and $R_7 \ge 0$.  Taking the integrated 
ratio we find:
\be \label{R7R4}
\left|{\int R_7 \over \int R_4}\right| = { 5 \int |F_{4}|^2   +  7 \int |F_{7}|^2  \over 4 \int |F_{4}|^2 + 8\int |F_{7}|^2} \le \frac54~.
\ee
This obviously does not satisfy the requirement \eqref{crazynumber} and we conclude that scale separation is not possible
in M-theory with only fluxes. Despite the result \eqref{R7R4}, many scale separated vacua with fluxes exist in M-theory, see for instance \cite{Becker:1996gj}, but crucially rely on higher derivative corrections, which we ignore. 

It is somewhat surprising that we find such a strong bound on the curvature scalars in M-theory,
given that constructions in massless type IIA with O6 planes uplift to purely geometric compactifications with 
fluxes in 11 dimensions. Although it is  counterintuitive, the no-gos of  \cite{Maldacena:2000mw, deWit:1986xg, Gibbons:1984kp} can also be applied to such constructions, ruling out Minkowski and de Sitter compactifications.  This is confirmed by the results of \cite{Hertzberg:2007wc}, where it was shown in 10 dimensions that O6 compactifications, with all fluxes turned on except the Romans mass, can never lead to de Sitter vacua or substantial amounts of inflation.  The results of this section can be applied to these compactifications, ruling out scale separation even in the presence of orientifolds.

\subsection{Scale separation in 10 dimensions}
In this section we treat type II supergravity in the democratic formalism:
\be \label{10daction}
S = M_{10}^8\int d^{10}X \sqrt{-g}  \lp R- {1\over2} (\partial \phi)^2 - { 1 \over 2} \e^{-\phi}|H_3|^2 - \frac14 \sum_n \e^{\frac{5-n}{2}\phi} 
|F_n|^2 \rp~,
\ee
where in the sum over $p$ we have
\be
n = \begin{cases} 1, \, 3, \, 5, \, 7, \,9 &\mbox{type IIB } \\ 
0,\, 2, \, 4, \, 6, \, 8, \, 10 & \mbox{type IIA,}  \end{cases} 
\ee
and $F_0$ is the Romans mass. In the democratic formulation the duality relation between RR field strengths must be imposed on-shell
\[
F_{10-n} = \e^{\frac{5-n}{2}\phi}(-1)^{\frac{(9-n)(8-n)}{2}}\star_{10} F_n~.
\]
We use this relation to treat all fluxes as internal, or magnetic. This is not strictly necessary, however 
in all cases one can dualize the result so that only magnetic fluxes appear.

For completeness we will also include the effects
 of localized D-branes. This amounts to adding 
\be\label{sourceaction}
S_\text{loc} = -N_p\mu_p\int_{\Sigma_{p+1}} \left\{\e^{\frac{p-3}{4}\phi} \sqrt{-P[g]} + (-1)^{\frac{p(p-1)}{2}} P[C_{p+1}]\right\}~,
\ee
to the action \eqref{10daction}, representing $N_p$ D-branes of charge $\mu_p$ coupled to the gauge potential $C_{p+1}$, given by:
\be
F_{p+2} = \dd C_{p+1} - H_3 \w C_{p-1}~.
\ee
The symbol $P[\cdots]$ denotes the pullback of the corresponding bulk fields onto the sub-manifold $\Sigma_{p+1}$ which the brane wraps. 
Note that in order to preserve Lorentz invariance, the brane must wrap the 4 spacetime dimensions and $p-3$ internal directions. Calculating the energy-momentum tensor of the matter fields and sources and plugging them into eqs.~\eqref{externalcurv} and \eqref{internalcurv} we find:
\be \label{r6r4nochmal}
\begin{split}
R_6 &= {1\over2} (\partial \phi)^2 +{3\over4} \e^{-\phi}|H_3|^2  + \sum_{n \le 6} {3+n\over 8} \e^{\frac{5-n}{2}\phi}|F_{n}|^2+\sum_{p\ge 3}\frac{15-p}{8}N_p\mu_p
\e^{\frac{p-3}{4}\phi}\delta(\Sigma_{p+1})~, \\
R_4 &= -{1\over2} \e^{-\phi}|H_3|^2 - \sum_{n \le 6}  {n-1\over 4} \e^{\frac{5-n}{2}\phi}|F_{n}|^2-\sum_{p \ge 3}\frac{7-p}{4}N_p\mu_p
\e^{\frac{p-3}{4}\phi}\delta(\Sigma_{p+1})~,
\end{split}
\ee
where $\delta(\Sigma_{p+1})$ is a Dirac delta function that localizes the D-brane contribution onto the brane worldvolume.  Again, we see that $R_6>0$ and $R_4<0$ as expected from the no-go theorems of \cite{Gibbons:1984kp,deWit:1986xg,Maldacena:2000mw}.

  To simplify the following equations we define
\be
\begin{split}
h &= \int \e^{-\phi}|H_3|^2~,\quad f_n = \int \e^{\frac{5-n}{2}\phi}|F_{n}|^2~,\\
d_p &= N_p\mu_p\int_{\Sigma_{p+1}}\e^{\frac{p-3}{4}\phi} \sqrt{-P[g]}~, \quad \sigma = \int (\partial \phi)^2~.
\end{split}
\ee
The integrated curvatures then reduce to
\be \label{r6r4fin}
\begin{split}
\int R_6 &= {1\over2}\sigma +{3\over4} h  + \sum_{n} {3+n\over 8} f_n + \sum_p\frac{15-p}{8} d_p~, \\
\int R_4 &= -{1\over2} h - \sum_{n}  {n-1\over 4} f_n - \sum_p\frac{7-p}{4} d_p~.
\end{split}
\ee

We will now measure scale separation, as before, by examining the ratio of the curvatures:
\be \label{scalesepII}
\left|{\int R_6 \over \int R_4}\right| =  {  {1\over2}\sigma +{3\over4} h  + \sum_{n} {3+n\over 8} f_n + \sum_p\frac{15-p}{8} d_p      \over   
{1\over2} h + \sum_{n}  {n-1\over 4} f_n + \sum_p\frac{7-p}{4} d_p  }.
\ee
This expression suggests a number of ways to construct a vacuum with a large hierarchy of scales: In type IIA one can seemingly
arrange for the Romans mass (in this equation $f_0$) to cancel the fluxes so as to obtain a small (or even positive) cosmological constant. This point was specifically 
addressed in \cite{Maldacena:2000mw} where it was shown that the equations of motion forbid such a cancellation. A more promising avenue is to 
exploit the fact that contributions from the dilaton, $\sigma$, the type IIB axion, $f_1$, and D7-branes, $d_7$, only appear in the numerator of eq. \eqref{scalesepII}. One can
then dial up these contributions in order to obtain a large ratio. This suggests that type IIB/F-theory compactifications could have 
an abundance of scale separated vacua, provided they can be stabilized with fluxes.

Using the dilaton equation of motion, we can further simplify this expression.
Going back to \eqref{10daction} and \eqref{sourceaction} we find the integrated dilaton equation of motion implies
\be \label{dilatonEOM}
{ 1 \over 2} h - \sum_n \frac{5-n}{4}f_n - \sum_p\frac{p-3}{4}d_p = 0~.
\ee
Using this to replace $h$ in \eqref{scalesepII} we find:
\be \label{scalesepIInoh}
\left|{\int R_6 \over \int R_4}\right| =  {  {1\over2}\sigma  + \sum_{n} {9-n\over 4} f_n + \sum_p\frac{3+p}{4} d_p      \over   
 \sum_{n}   f_n + \sum_p d_p  }~.
\ee
We now see that the only possibility for a hierarchy of scales is for 
\be \label{resultscalesep}
\sigma \gg \sum_{n}   f_n + \sum_p d_p~.
\ee
Otherwise, if the dilaton gradients can be neglected, \eqref{scalesepIInoh} is bounded by three\footnote{This conclusion can be reached in analogy with centre of mass computations; a weighted average, $r_{\rm com} = \sum r_i m_i/ \sum m_j$,  is bounded by the largest contribution, $r_i$.}.  

\subsection{Alternative perspective: integration with dilaton weights}

A complementary means reaching the conclusion above is to integrate all quantities with some non-trivial weighting.  Rather than integrating the curvatures (and the related equations of motion) solely with respect to the metric volume form $\sqrt{-g}$, one can  integrate  with respect to $e^{a\phi}\sqrt{-g}$, where the constant $a$ is chosen such that the combination $e^{a\phi}\sqrt{-g}\, g^{MN}$ is invariant under the constant rescaling
\be \label{constrescaling}
g_{MN} \to s\,g_{MN} \,,\quad e^{-\phi} \to s^2\,e^{-\phi} \, \quad F_n \to s^{{3-n \over 4}} F_n.
\ee
This rescaling is a symmetry of the classical equations of motion in the absence of sources, and it is the combination $\sqrt{-g}\, g^{MN}$ which is relevant in determining how the integrated curvatures $\int R_4$ and $\int R_6$ scale under it. Invariance of the integrated curvatures dictates that $a=2$ is the appropriate choice.

Introducing a superscript `$(\phi)$' to denote the weighting, {\it i.e.}
\be
\begin{split}
h^{(\phi)} &= \int \e^{\phi}|H_3|^2~,\quad f^{(\phi)}_n = \int \e^{\frac{9-n}{2}\phi}|F_{n}|^2~,\\
d^{(\phi)}_p &= N_p\mu_p\int_{N_{p+1}}\e^{\frac{p+5}{4}\phi} \sqrt{-P[g]}~, \quad \sigma^{(\phi)} = \int e^{2\phi}(\partial \phi)^2~.
\end{split}
\ee
we find that the integrated Einstein and dilaton equations of motion become
\be
\begin{split}
\int e^{2\phi}\, R_6 &= {1\over2}\sigma^{(\phi)} +{3\over4} h^{(\phi)}  + \sum_{n} {3+n\over 8} f^{(\phi)}_n + \sum_p\frac{15-p}{8} d^{(\phi)}_p~, \\
\int e^{2\phi}\, R_4 &= -{1\over2} h^{(\phi)} - \sum_{n}  {n-1\over 4} f^{(\phi)}_n - \sum_p\frac{7-p}{4} d^{(\phi)}_p
\end{split}
\ee
and
\be
-\sigma^{(\phi)}+{ 1 \over 2} h^{(\phi)} - \sum_n \frac{5-n}{4}f^{(\phi)}_n - \sum_p\frac{p-3}{4}d^{(\phi)}_p = 0~,
\ee
respectively. (The $\sigma^{(\phi)}$ term in the latter arises from integrating the $\square \phi$ term by parts.) Eliminating $h^{(\phi)}$ using the dilaton equation as before, we find that the ratio of curvatures becomes
\be \label{dilweightedratio}
\left|{\int e^{2\phi}\,R_6 \over \int e^{2\phi}\,R_4}\right| =  {  2\sigma^{(\phi)}  + \sum_{n} {9-n\over 4} f^{(\phi)}_n + \sum_p\frac{3+p}{4} d^{(\phi)}_p      \over   
\sigma^{(\phi)}+ \sum_{n}   f^{(\phi)}_n + \sum_p d^{(\phi)}_p  }~.
\ee
Therefore, integrating with respect to a weight that is invariant under the constant rescaling \eqref{constrescaling} ensures that the resulting integrated curvature ratio is always bounded by three (thus precluding scale-separated solutions). Nevertheless, it remains true that, if the dilaton has some wildly-varying profile in the internal dimensions, the ratio obtained in \eqref{dilweightedratio} will cease to be an appropriate estimate of the degree of scale separation. It is in this sense that this alternative approach agrees with the one used in deriving eq.~\eqref{resultscalesep}.

\section{An application: O6 singularities in massive type IIA} \label{sec:O6}
In this section we demonstrate that our findings provide concrete insights for type IIA flux compactifications. It has been shown in \cite{DeWolfe:2005uu} (see also \cite{Behrndt:2004km, Derendinger:2004jn, Acharya:2006ne, Caviezel:2008ik}), that O6 compactifications of \emph{massive} type IIA supergravity can have the following striking features: first, full moduli stabilisation at the classical level and second, an AdS vacuum where a parametric separation of scales simultaneously suppresses derivative and loop corrections. The parameter that allows for tuning is the flux quantum number, $n$, of the RR  4-form flux, $F_4$, that is not constrained by any tadpole condition. In particular one finds the following dependence of the string coupling, KK scale and cosmological constant on $n$:
\be
e^{\phi}\sim \frac{1}{n^{3/4}}\,,\qquad {M^2_{\rm KK} \over M^2_{\rm Pl}} \sim {1 \over n^{7/2}}\,,\qquad {M^2_{\Lambda} \over M_{\rm Pl}^2}\sim {1\over n^{9/2}}\,. 
\ee
This good news comes at a cost: the backreaction of the O6-planes is only captured in the limit where they are smeared over the compact dimensions. The validity of this approximation has been questioned in \cite{McOrist:2012yc}. Unlike the standard type IIB compactifications to 4-dimensional Minkowski space \cite{Giddings:2001yu, Dasgupta:1999ss}, the backreaction has not been computed since the compactification involves four intersecting O6-planes. Even in flat space, supergravity solutions describing 4 mutually orthogonal localised branes/planes are not known.

In reference \cite{Saracco:2012wc} it was demonstrated that the solutions \cite{DeWolfe:2005uu,Behrndt:2004km, Derendinger:2004jn, Acharya:2006ne, Caviezel:2008ik} become full-blown dynamic $\SU(2)$ structures when the O6-planes are localised.  Although the dynamic $\SU(2)$ structure implies that the solution is hopelessly complicated, reference \cite{Saracco:2012wc} provides some reason for optimism: despite the presence of O6-planes, the fully back-reacted metric can be regular.  This is achieved at the 10-dimensional level when the Roman's mass is non-zero, contrary to the usual massless resolution via an uplift to pure geometry in 11 dimensions.

\begin{figure}[h!]
		\begin{center}
		 \includegraphics[width=.75\textwidth]{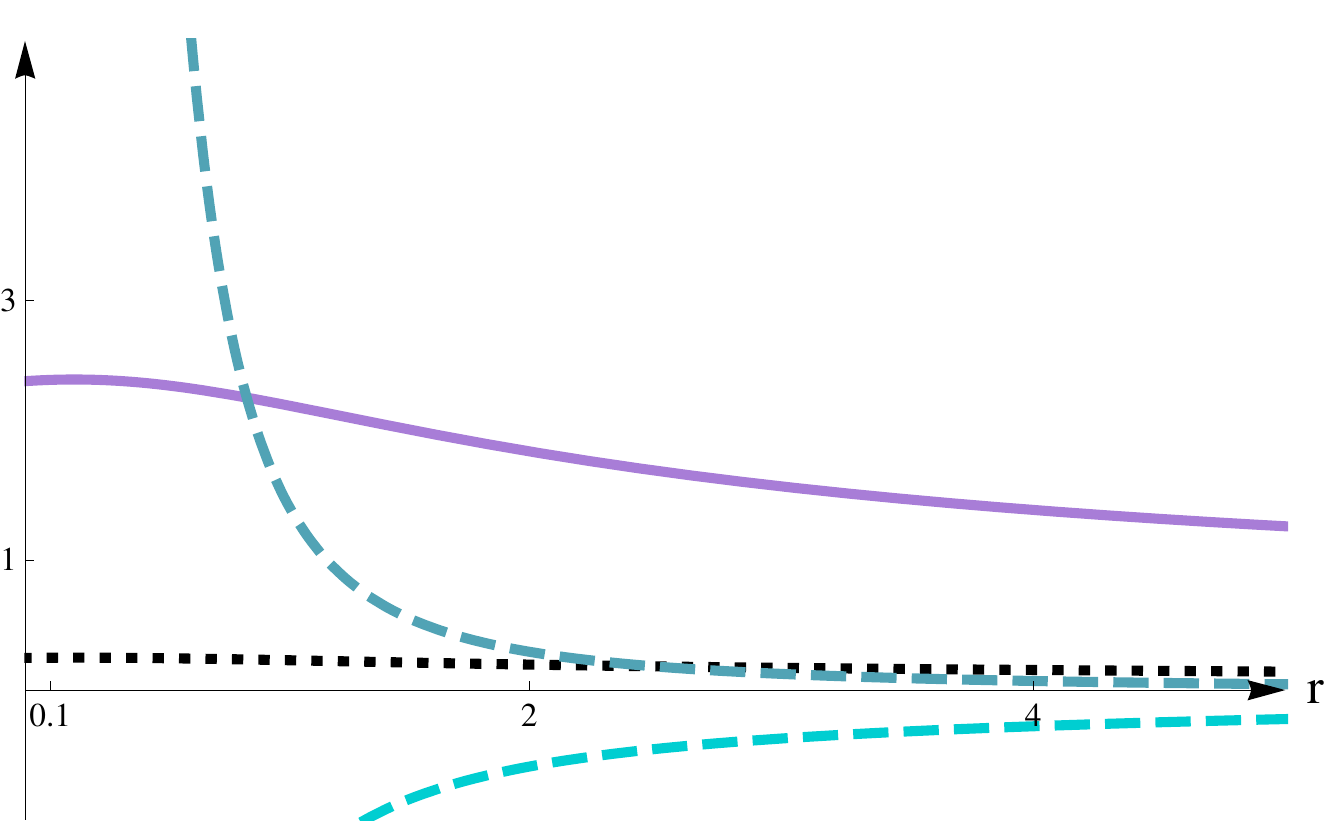}
		 \caption{\small{Figure 1b from \cite{Saracco:2012wc}. The horizontal axis denotes the distance from the O6 source when the solution is locally approximated to be spherically symmetric. The dotted line corresponds to the string coupling and is practically flat. The solid line is the warpfactor and indeed stays finite, indicating a smooth geometry.}}
		 \label{fig:o6m}
		 \end{center}	
		\end{figure}

We now argue that this resolution is unlikely to occur when back-reacting the O6-planes of \cite{DeWolfe:2005uu}. We do not claim that the resolution mechanism of \cite{Saracco:2012wc} is invalid\footnote{In fact some extra evidence for the existence of the resolution mechanism was found from a field theory viewpoint \cite{Saracco:2013aoa}.}, instead, we claim it does not occur in vacuum solutions with scale separation. Our reasoning is rather simple. If the geometry is smooth in 10 dimensions, then the bulk field equations are satisfied without the need for explicit source terms. But without sources --- particularly negative tension sources --- our results indicate that scale separation requires large dilaton variation, a feature absent in the resolved O6 solutions of \cite{Saracco:2012wc}. For clarity we have added Figure \ref{fig:o6m} of \cite{Saracco:2012wc} showing the profile of the dilaton\footnote{We are grateful to the authors of  \cite{Saracco:2012wc} for allowing us to reproduce this figure}.  From the figure one concludes that the dilaton is nearly constant.  The authors of \cite{Saracco:2012wc} also solve the system perturbatively around the massless O6 solution and find that the dilaton is constant at first order. 

\section{Discussion}

The purpose of this note is to establish conditions for constructing genuinely lower-dimensional vacua. Our calculation  applies to 10- and 11-dimensional supergravity with fluxes, excluding higher-derivative corrections. We find that in the absence of the negative tension induced by orientifold singularities, the level of scale separation is controlled by dilaton gradients in 10-dimensional supergravity.  In 11-dimensional supergravity we show that scale separation is impossible.  This shows that the usual no-go theorems of \cite{Gibbons:1984kp, deWit:1986xg, Maldacena:2000mw} against Minkowski and de Sitter vacua from fluxes extend to AdS vacua with small values of the cosmological constant compared to the KK scale. This provides a natural explanation for the observation in \cite{Tsimpis:2012tu} that a  class of warped AdS vacua has no  solutions which exhibit scale separation.  We also  demonstrate that the classical resolution of O6 singularities due to Romans mass \cite{Saracco:2012wc, Saracco:2013aoa} cannot take place in a compactification with scale separation such as those of \cite{DeWolfe:2005uu}.  This is because these constructions do not exhibit the large dilaton gradients that provide the only hope for scale separation in constructions without negative tension sources. Finally, our results have immediate applications for model building.

As discussed in the introduction, any scale separated vacuum in a theory with extra dimensions will have a cosmological hierarchy problem. This statement has implications for the string landscape.
In a theory with a landscape of vacua, many apparent fine-tuning problems are solved by the anthropic principle. To avoid any use of anthropic reasoning, either the landscape or a mechanism to populate it, \emph{e.g.}~eternal inflation, should be absent from the theory.  As observers in a particular vacuum, the difficulty lies in identifying when to apply anthropic reasoning.   It is widely accepted that the cosmological constant problem can be solved by anthropic reasoning \cite{Weinberg:1988cp}, whereas the number of macroscopic spacetime dimensions might have a dynamical explanation \cite{Brandenberger:1988aj}. Given the relation between the cosmological hierarchy and scale separation, these viewpoints are inconsistent.  So either we have to accept that there is a dynamical mechanism for selecting 4 large spacetime dimensions which implies a cosmological hierarchy (but softens it significantly), or the reason we observe 4 dimensions is anthropic. The first possibility is clearly the more exciting option.

It is natural to think about scale separated AdS vacua from a holographic perspective. In particular the questions, ``When is the cosmological constant small with respect to the KK scale?'' and ``Is there a landscape of such solutions?'' should have a clear translation in the dual CFT language. Some general properties of CFTs dual to scale separated AdS vacua are known \cite{Polchinski:2009ch, Papadodimas:2011kn, deAlwis:2014wia}, \emph{e.g.} they would exhibit a parametrically large gap in the spectrum of dimensions of conformal operators. Although studying scale separation from the holographic viewpoint would be of great interest, no explicit CFT example is known.

\vspace{8pt}

\section*{Acknowledgements}

We are happy to acknowledge useful discussions with N. Bob{\v e}v, M. Kleban, S. Sethi and A. Tomassielo. TVR furthermore likes to thank the organisers of the workshops ``Fine-Tuning, Anthropics and the String Landscape'' (Madrid 2014) and the ``Gordon Research Conference: String Theory $\&$ Cosmology'' (Hong Kong 2015) for providing the right environment for thinking about the matters discussed in this paper.   This work is supported by the National Science Foundation of Belgium (FWO) grant G.0.E52.14N Odysseus and postdoctoral fellowship programmes, and the European Union's Horizon 2020 research and innovation programme under the Marie Sk{\l}odowska-Curie grant agreement No 656491. We acknowledge support from the European Science Foundation HoloGrav Network, the Belgian Federal Science Policy Office through the Inter-University Attraction Pole P7/37, and the COST Action MP1210 `The String Theory Universe'.

{\footnotesize
	\bibliography{refs}} 
\bibliographystyle{utphysmodb}

\end{document}